\documentclass[
aps, prb,
superscriptaddress,
twocolumn,
amsmath,amssymb
]{revtex4-2}

\usepackage{chemformula} 
\usepackage{siunitx}
\usepackage{graphicx}
\usepackage{subcaption}
\usepackage[percent]{overpic}
\usepackage{subfiles}
\usepackage[hidelinks]{hyperref}
\usepackage[capitalise]{cleveref}

\def \ESA {\ch{EuSn2As2}}
\def \MBT {\ch{MnBi2Te4}}

\def \IFW {Leibniz Institute for Solid State and Materials Research,
IFW Dresden, Helmholtzstrasse 20, 01069 Dresden, Germany}
\def \FIAN {V. L. Ginzburg Research Center at P. N. Lebedev Physical Institute, RAS, Moscow, Russia}
\def \ctqmat {W\"{u}rzburg-Dresden Cluster of Excellence ctd.qmat, 01062 Dresden, Germany}
\def \TUD {Department of Physics, TU Dresden, D-01062 Dresden, Germany}
\def \CNRST {CNRS, Laboratoire National des Champs Magnétiques Intenses, Université Grenoble-Alpes, Université Toulouse 3, INSA-Toulouse, EMFL, 31400 Toulouse, France}
\def \CNRSG {Université Grenoble Alpes, CNRS, CEA, Grenoble-INP, Spintec, F-38000 Grenoble, France}

\begin{document}

\preprint{APS/123-QED}

\title{Anomalous and Topological Hall Effects in Antiferromagnetic \ESA\ Nanostructures}

\author{E.~I. Maltsev}
    \email{e.maltsev@ifw-dresden.de}
    \affiliation{\IFW}
    \affiliation{\ctqmat}
\author{N. P\'{e}rez}
    \affiliation{\IFW}
\author{R. Giraud}
    \affiliation{\IFW}
    \affiliation{\CNRSG}
\author{K.~K. Bestha}
    \affiliation{\IFW}
\author{A.~U.~B. Wolter}
    \affiliation{\IFW}
\author{J. Dufouleur}
    \affiliation{\IFW}
    \affiliation{\ctqmat}
\author{K.~S. Pervakov}
    \affiliation{\FIAN}
\author{V.~M. Pudalov}
    \affiliation{\FIAN}
\author{K. Nielsch}
    \affiliation{\IFW}
\author{B. B\"{u}chner}
    \affiliation{\IFW}
    \affiliation{\ctqmat}
    \affiliation{\TUD}
\author{L. Veyrat}
    \affiliation{\IFW}
    \affiliation{\ctqmat}
    \affiliation{\CNRST}

\date{\today}

\begin{abstract}
We investigate magnetotransport in exfoliated nanostructures of the candidate magnetic 3D topological insulator \ESA.
Similar to macroscopic single crystals, the negative magnetoresistance observed below the N\'eel temperature ($T_N$ = \qty{24}{\kelvin}) is related to the canted antiferromagnetic state (CAF) of an easy-plane antiferromagnet (AFM), with an increase of the saturation field when tilting the applied magnetic field away from the (ab) plane ($\mu_{0} H^{c}_{s}$ = \qty{4.9}{\tesla}, $\mu_{0} H^{ab}_{s}$ = \qty{3.6}{\tesla}). 
Higher-accuracy measurements in nanostructures up to \qty{14}{\tesla} further evidence a non-linear normal Hall response due to several electronic bands.
Interestingly, the transverse resistance due to magnetism reveals an anomalous Hall effect in the CAF state, but also a topological Hall effect due to chiral spin textures, as found in AFM or helical magnets.
The presence of real-space chiral spin texture, already reported in another magnetic topological insulator, \MBT, could be a characteristic generally appearing in magnetic 3D topological insulators.
\end{abstract}


\maketitle

\section{Introduction}

Intrinsically magnetic topological materials have recently attracted strong interest due to topological phases controlled by magnetic order, such as the axion and the quantum anomalous Hall insulators \cite{Li_2019, Li_2019a}.
In these systems, magnetism, electronic and topological properties are strongly connected: while magnetism is responsible for the emergence of topological phases (opening a gap in the band structure of topological surface states), it also strongly impacts the classical electronic properties.
In magnets generally, the presence of magnetic textures can impact both the longitudinal magnetoresistance \cite{Viret_1996, Gregg_1996} and the Hall effect \cite{Xia_2024, Schlitz_2019, Kimbell_2022}.
Specifically, chiral magnetic textures give rise to an additional contribution to the Hall effect not proportional to the magnetization.
This Hall response arises from a Berry phase induced by scattering on local real-space chiral spin textures (such as skyrmions \cite{Kimbell_2022,Schlitz_2019} or chiral domain walls \cite{Xia_2024}).
However, this so-called topological Hall effect (THE) is not necessarily related to the non-trivial topology of magnetic textures in real space.
It is actually the specific response of chiral spin textures, not limited to skyrmions \cite{Kimbell_2022, Yi_2024}, and could therefore be referred to as "spin chirality Hall effect".
However, we will use the term THE in this work for consistency with the literature, without implying the topological nature of the underlying cause of the effect.

\begin{figure}[t]
    \centering
    \includegraphics[width=0.4\textwidth]{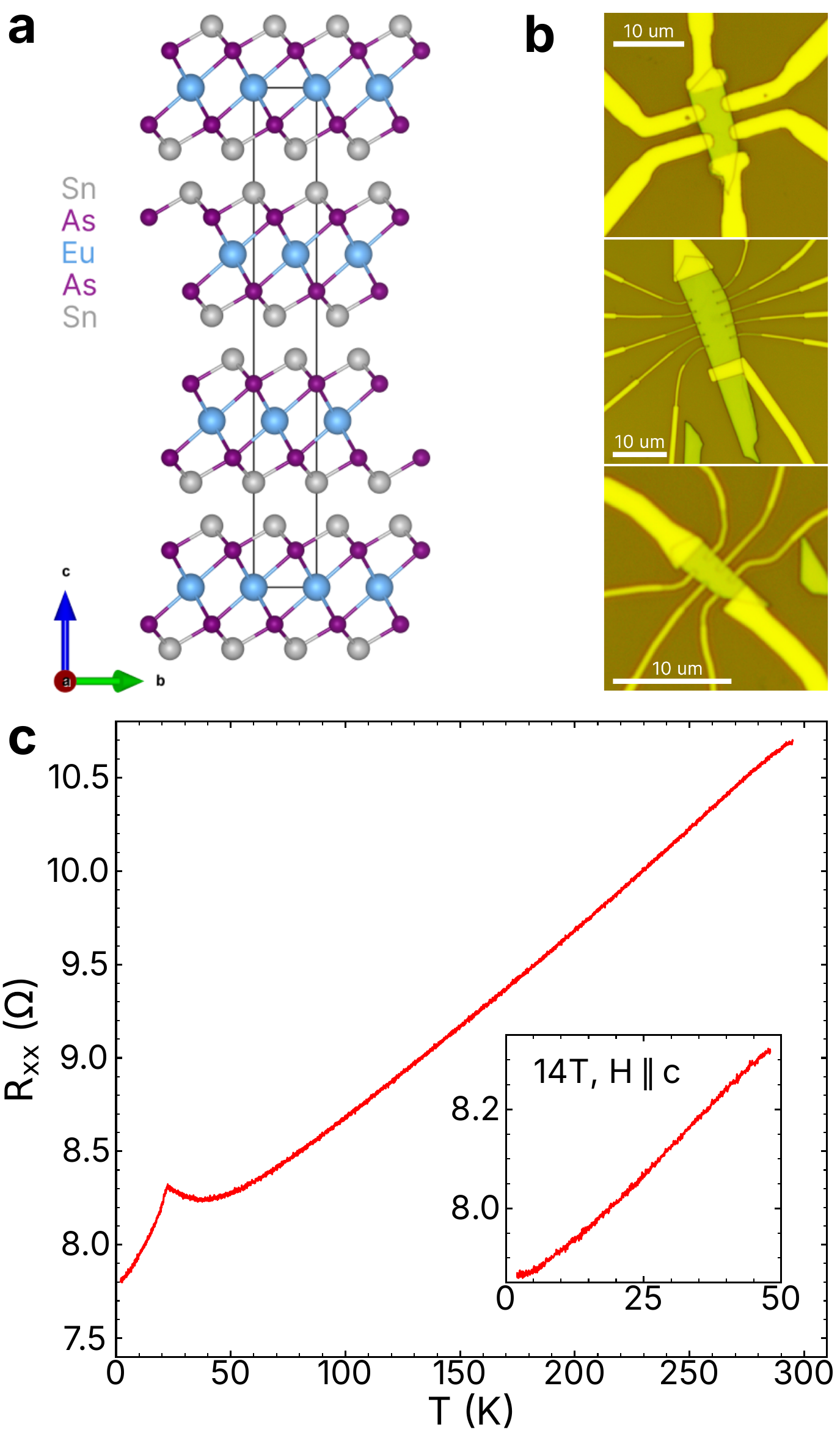}
    \begin{subcaptiongroup}
        \phantomcaption\label{subfig:crystal_structure}
        \phantomcaption\label{subfig:nanostructures}
        \phantomcaption\label{subfig:R_vs_T}
    \end{subcaptiongroup}
    \caption{\label{fig:one} (\subref{subfig:crystal_structure}) Crystal structure of \ESA; (\subref{subfig:nanostructures}) studied nanostructures: \qtylist{140; 110; 60}{\nm} (from top to bottom); (\subref{subfig:R_vs_T}) temperature dependence of longitudinal resistance $R_{xx}$ of \qty{110}{\nm} sample at zero magnetic field and \qty{14}{\tesla} (inset).}
\end{figure}

In magnetic topological insulators, it is particularly important to understand the transport properties of magnetic textures, such as domain walls, and their possible contribution to the Hall response.
For instance, the presence of domain walls may explain the deviation from perfect quantization of the transverse resistance in the quantum anomalous Hall regime of \MBT\ \cite{Bai2023}.
Also, several recently discovered magnetic 3D topological insulators (3DTIs) are antiferromagnets \cite{Otrokov_2019, Li_2019, Takeda_2024, Yan_2022}, possibly hosting chiral domain walls or helical states.
Among them, \ESA\ is a layered van der Waals (vdW) compound with the same crystal structure as \MBT\ ($R\bar{3}m$, see \cref{subfig:crystal_structure}) \cite{Arguilla_2017}.
According to DFT calculations, \ESA\ is a strong topological insulator in the paramagnetic phase (PM) and an axion insulator in the antiferromagnetic phase (AFM), with gapless topological surface states (TSS) \cite{Li_2019}.
However, ab-initio calculations also predict \ESA\ to be heavily p-doped ($\approx$ \qty{0.25}{\eV}), so that TSS are not filled and charge transport is dominated by valence-band bulk carriers.
This was confirmed by both standard and time-resolved ARPES measurements, which agree well with the DFT calculated bands \cite{Li_2019, Lv_2022}.
Below $T_N$ = \qty{24}{\kelvin}, \ESA\ orders in an A-type antiferromagnetic state (ferromagnetic vdW layers are coupled antiferromagnetically) with an easy-plane anisotropy, as revealed by bulk magnetisation measurements, with a small anisotropy energy of \qty{1.4e5}{\J\per\cubic\m} \cite{Golovchanskiy_2022}.
However, several similar materials (antiferromagnetic 3DTI like \MBT\ or layered AFM like \ch{EuIn2As2} \cite{Lee_2019, Yan_2022}) additionally present chiral magnetic textures, which can be evidenced by a THE.
It is reasonable that such features could also be present in \ESA.
However, no evidence of such textures were reported so far.

In this work, we investigate the electronic and magnetic properties of \ESA\ through magnetotransport experiments on exfoliated nanostructures with thicknesses varying between \qtylist{140; 60}{\nm} (exhibiting bulk-like properties).
The magnetoresistance measured is compatible with that already reported for macroscopic samples, with a negative magnetoresistance in the canted-antiferromagnetic (CAF) phase and a larger saturation field $\mu_{0} H^{c}_{s}\approx\ $\qty{4.9}{\tesla} for a magnetic field applied perpendicular to the (ab) plane \cite{Arguilla_2017, Golovchanskiy_2022, Li_2019, Chen_2020, Pakhira_2021}.
The angular dependence of the saturation field is also compatible with an easy-plane AFM structure, with an anisotropy field much smaller than the inter-layer exchange field.

Moreover, our measurements of the transverse resistance studied up to \qty{14}{\tesla} reveal several additional features not reported before.
First, we observe a non-linear normal Hall effect, probably due to multi-band transport.
Second, a careful analysis of the Hall response due to magnetism reveals two contributions.
In addition to the anomalous Hall effect, we give clear evidence, for the first time, of a topological Hall effect in the CAF regime.
This implies the existence of chiral magnetic textures in \ESA, a property shared by other magnetic 3DTIs such as \MBT\ \cite{Lee_2019}.

\section{Results and discussion}

We investigated the magneto-transport properties of three \ESA\ thin flakes with thicknesses \qtylist{140; 110; 60}{\nm}.
All flakes gave qualitatively similar results, so that the following discussion focuses on the \qty{110}{\nm} flake, studied most extensively.
The results on the other two flakes are presented in the Supporting Information (SI).
As seen in \cref{subfig:R_vs_T}, the temperature dependence of the longitudinal resistance shows a metallic behavior with a peak at \qty{24}{\kelvin}, as previously observed in the \ESA\ compound \cite{Arguilla_2017, Chen_2020, Li_2021, Pakhira_2021}.
This resistance peak is typical for the onset of magnetic fluctuations close to a phase transition, here the antiferromagnetic transition, followed by a more rapid decrease of the resistance with temperature when long-range magnetic order is established.
As seen in the inset of \cref{subfig:R_vs_T} the peak vanishes when applying a magnetic field of \qty{14}{\tesla}, as magnetic fluctuations are suppressed.
The residual resistivity ratio (RRR) value equals 1.4 (consistent with other reports \cite{Arguilla_2017, Chen_2020, Pakhira_2021}), implying that the material is strongly disordered.
The following values of resistivity were measured at low temperatures (\qtyrange{2}{3}{\kelvin}): \qtylist{301; 292; 367}{\micro\ohm\cm} for \qtylist{140; 110; 60}{\nm} flakes, respectively.
Recalculating resistivities to conductivities gives the following values: \qtylist{3.3; 3.4; 2.7}{\kilo\siemens\per\cm}.
These small differences from sample to sample are partially induced by the invasive contacts deposited on the flakes, which affect the flow of electrons in the nanostructure and can lead to reduced measured longitudinal and Hall voltages \cite{Gluschke_2020}.

\begin{figure*}
    \centering
    \includegraphics[width=\textwidth]{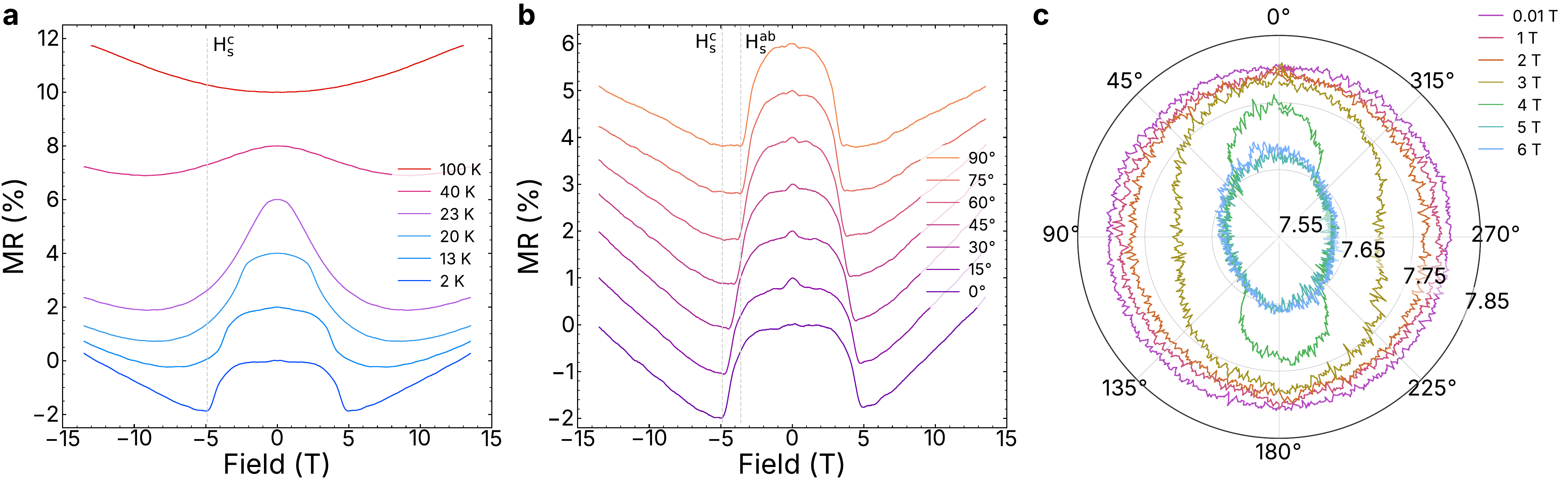}
    \begin{subcaptiongroup}
        \phantomcaption\label{subfig:MR_OOP_110nm}
        \phantomcaption\label{subfig:MR_tilts_110nm}
        \phantomcaption\label{subfig:Rxx_polar_110nm}
    \end{subcaptiongroup}
    \caption{\label{fig:MR_110nm} Magnetoresistance of \ESA\ \qty{110}{\nm} flake (\subref{subfig:MR_OOP_110nm}) for different temperatures in an out-of-plane orientation (symmetrized data to account for contacts misalignment, vertical shift of 2 \% is added for clarity), (\subref{subfig:MR_tilts_110nm}) for different field orientations at \qty{2}{\kelvin} (data without symmetrizaion, vertical shift of 1 \% is added for clarity).
    (\subref{subfig:Rxx_polar_110nm}) Polar diagram of $R_{xx}$ for different applied magnetic field strength.
    \ang{0} is out-of-plane, \ang{90} is in-plane.}
\end{figure*}

Next we investigate the magnetotransport properties of \ESA\ nanostructures.
We first measured the dependence of the resistance with respect to magnetic field (magnetoresistance, MR) for the out-of-plane (OOP) field orientation, at several temperatures, as shown in \cref{subfig:MR_OOP_110nm}. 
At temperatures higher than \qty{60}{\kelvin}, the quadratic positive MR is only due to cyclotron orbits and its amplitude is directly related to the carriers' mobility.
At temperatures lower than \qty{60}{\kelvin}, magnetic correlations develop in the material and give rise in addition to a negative contribution to MR.
Moreover, below $T_{N}$, another negative MR component develops, with a clear dome shape below the saturation field $H_{s}$ (where magnetic moments are fully aligned along the field direction).
A recent theoretical proposal suggests to explain this dome-shape negative MR as a consequence of the violation of Z2 symmetry and the different localization of electron wave functions on either AFM sublattices depending on the electron spin \cite{Grigoriev_2025}.
This theory is supported by the isotropy of this MR shape with field direction (see \cref{subfig:MR_tilts_110nm}).

We then investigate the effect of tilting the magnetic field between the out-of-plane ($\theta=$~\qty{0}{\degree}, along \textit{c}-axis) and the in-plane ($\theta=$~\qty{90}{\degree}, in the \textit{ab}-plane) orientations (see \cref{subfig:MR_tilts_110nm}).
The $H_{s}$ value depends on temperature and on the field orientation, with $H_{s}^{ab} < H_{s}^{c}$ for an easy-plane AFM.
This is confirmed by the MR measured at different angles between the OOP and the in-plane (IP) configuration, at \qty{2}{\kelvin}, as shown in \cref{subfig:MR_tilts_110nm}.
MR measured in OOP and IP configurations are very similar which is compatible with the (ab) easy-plane with small anisotropy (with respect to exchange coupling) reported in \ESA.
The only qualitative difference between OOP and IP is the appearance of a peak in the low-field region, observed on a wide angular range but not for the OOP configuration.
This narrow feature is better visible in Fig. S7, which focuses on the low field region, \qtyrange[range-phrase=\ to\ ]{-1}{1}{\tesla}. 
This peak is not compatible with standard weak-localization theory, as it does not appear with out-of-plane field but only with in-plane magnetic field, for which the coherent loops occur in the thickness of the nanostructure and are expected to be smaller.
However, the field range of the peak ($\approx$ \qty{0.12}{\tesla}) corresponds to the field required to align antiferromagnetic domains perpendicular to the external field $H^{ab}$ \cite{Pakhira_2021, Pakhira_arXiv2024}.
The existence of such domains was confirmed independently with measurements of the rotational anisotropy of optical second-harmonic generation (RA-SHG) \cite{Saatjian_2025}.

\begin{figure*}
    \centering
    \includegraphics[width=0.9\textwidth]{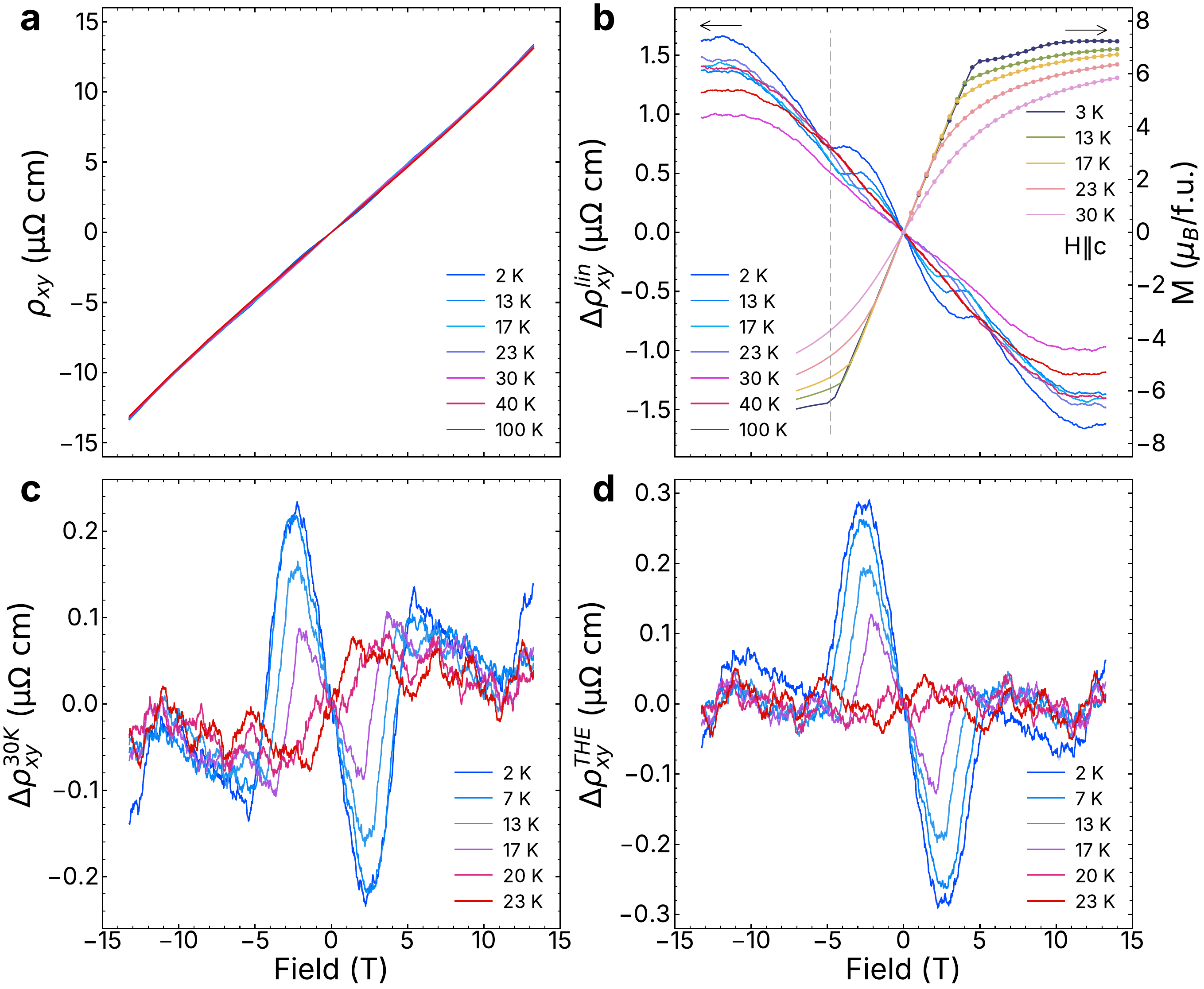}
    \begin{subcaptiongroup}
        \phantomcaption\label{subfig:Hall_110nm}
        \phantomcaption\label{subfig:AHElin_110nm}
        \phantomcaption\label{subfig:AHE_110nm}
        \phantomcaption\label{subfig:THE_110nm}
    \end{subcaptiongroup}
    \caption{\label{fig:Hall_110nm} Hall measurements on the \qty{110}{\nm} flake. (\subref{subfig:Hall_110nm}) Antisymmetrized Hall effect measured at different temperatures.
    (\subref{subfig:AHElin_110nm}) Residuals of the Hall effect in \textit{a} after subtracting a high-field (11-13T) linear slope.
    Magnetization curves measured on a single crystal (scatter points -- measured, solid lines -- interpolation) are shown for reference.
    Below saturation fields an anomaly is clearly visible for $T < T_N$.
    (\subref{subfig:AHE_110nm}) Residual Hall response after suppression of the normal Hall contributions calculated as $\rho_{xy}(T) - \rho_{xy}(\mathrm{30K})$, showing two contributions to the Hall effect from magnetism.
    (\subref{subfig:THE_110nm}) Topological Hall effect: residuals of Hall effect curves after subtraction of $\rho_{xy}(\mathrm{30K})$ and of the scaled magnetization data ($\rho_{xy}(T) - \rho_{xy}(\mathrm{30K}) - \alpha \Delta M(T)$, see main text).}
\end{figure*}

From the MR curves of \cref{subfig:MR_tilts_110nm}, we determine the evolution of the saturation fields from OOP ($\mu_{0} H_{s}^{c}$ = \qty{4.9}{\tesla}) to IP ($\mu_{0} H_{s}^{ab}$ = \qty{3.6}{\tesla}) configuration corresponding to the field boundary of the dome-shape negative MR.
The correspondence between the end of the dome-shaped negative MR and the saturation field is confirmed by magnetisation measurements performed on bulk crystals (see below) and by other studies \cite{Arguilla_2017,Li_2019,Chen_2020,Pakhira_2021,Golovchanskiy_2022, Grigoriev_2025}.
Next, we calculate exchange and anisotropy energies $J$ and $K_{u}$ following the analysis in \cite{Golovchanskiy_2022}. For that, we use the following relations:
\begin{align*}
& \mu_{0} H_{s}^{ab} = 2 \mu_{0} H_{e} = 2J/M_{s} \\
& \mu_{0} H_{s}^{c} = 2 \mu_{0} H_{e} + \mu_{0} M_{s} + 2 K_{u}/M_{s} \\
& \mu_{0} M_{eff} = \mu_{0} H_{s}^{c} - \mu_{0} H_{s}^{ab}
\end{align*}
and hence $\mu_{0} M_{eff} = \mu_{0} M_{s} + 2 K_{u}/M_{s}$, where $M_{s}$ is the saturation magnetization extracted from magnetization measurements on single crystals (\cref{subfig:AHElin_110nm}) and $H_{e}$ is the exchange field.
The value $\mu_{0} M_{s}$ = \qty{0.62}{\tesla} is obtained as $\mu_{0} \mu_{B} z M_{s}/ V_{cell}$ where $z$ is the number of Eu atoms per unit cell and $V_{cell}$ is the unit cell volume.
The results are $J$ = \qty{8.9e5}{\joule\per\cubic\meter}, $K_{u}$ = \qty{1.7e5}{\joule\per\cubic\meter}, and $H_{e}$ = \qty{14e5}{\joule\per\cubic\meter} ($\mu_{0}H_{e}$ = \qty{1.8}{\tesla}).
The small discrepancy with the values found in \cite{Golovchanskiy_2022} ($J$ = \qty{5.7e5}{\joule\per\cubic\meter} and $K_{u}$ = \qty{1.4e5}{\joule\per\cubic\meter}) could be explained by the small size of the crystal we used for magnetization measurements (\qty{0.19 \pm 0.02}{\mg}), which induces an uncertainty of \qty{11}{\percent} on the final magnetization amplitude $M_{s}$.

We also measured MR in IP orientation at different temperatures (see Fig. S5).
The temperature dependence of MR in OOP and IP orientations is very similar, since the evolution of the field-induced CAF state is mostly related to the inter-layer exchange coupling.
The low-field peak decreases and smears with increasing temperature, and disappears at $T_{N}$ implying a relation to magnetic ordering.

Finally, in order to better evidence the magnetic anisotropy, the longitudinal resistance was measured with respect to the tilt angle at constant magnetic fields (\cref{subfig:Rxx_polar_110nm}).
For fields lower than \qty{1}{\tesla}, the resistance appears mostly isotropic (nearly circular traces in \cref{subfig:Rxx_polar_110nm}), as the magnetic moments mainly remain in the \textit{ab}-plane at low field. Between \qty{1}{\tesla} and $\mu_{0} H_{s}^{ab} \simeq$ \qty{3.6}{\tesla}, where the dome-shaped negative magnetoresistance manifests, the MR becomes more strongly anisotropic, with lower in-plane MR, as can be seen by the ovoid traces.
This MR anisotropy continues even above $\mu_{0} H_{s}^{c} \simeq$ \qty{4.9}{\tesla}. 
Between $H_{s}^{ab}$ and $H_{s}^{c}$, the saturation field is reached at intermediate angles, resulting in an abrupt jump of the MR in the \qty{4}{\tesla} trace.

Comparison of our data and previous reports reveals several discrepancies.
First, the maximal magnitude of negative MR is different in all measurements. Our results (-4 \% for OOP and -5 \% for IP) are slightly lower (but within our own sample-to-sample variation) than was observed in \cite{Li_2021}, and two times smaller than in \cite{Chen_2020}.
It is similar to values recently reported on nanostructures \cite{Ma_2026}.
Second, we do not confirm a kink in MR related to a second magnetization plateau which was observed in \cite{Li_2021}.
However, the kink in the magnetisation was observed in bulk crystals (see \cref{subfig:AHElin_110nm}).
We expect this kink to be due to a secondary phase present in the bulk crystals, but absent in the mechanically exfoliated flakes.
Third, the maximal magnitude of negative MR is measured at $T_{N}$ in our samples which is in contrast to \cite{Chen_2020, Li_2021} where it was observed at their lowest temperature.
This observation is however reproducible for two of our samples in both OOP and IP orientations (see SI).
Fourth, we confirm the observation of a zero-field peak when tilting the field toward IP orientation, as observed in \cite{Li_2021}.
But we do not confirm the zero-field peak in OOP orientation.

\section{Hall effect}

Next we turn to the central point of our study: the transverse resistance with OOP field orientation, which was measured simultaneously with longitudinal MR (\cref{subfig:Hall_110nm}).
We observe that the Hall effect is almost temperature independent with hole-type charge carriers dominating the transport.
That agrees with bulk measurements \cite{Chen_2020, Li_2021} and ARPES studies \cite{Li_2019}.
However, contrary to the bulk measurements, our curves for exfoliated flakes are not linear and have a visible S-shape.
This S-shape with an upturn at high fields could indicate multi-band transport, with a minority electron carrier on top of the majority hole carriers, as was observed in ARPES \cite{Li_2019}.
Furthermore, we observe contributions to the Hall effect related to magnetism at low temperatures.
In addition to the anomalous Hall effect (AHE) related to the net magnetization induced by external field, we evidence a topological Hall effect related to chiral textures.
In the following, we will first analyze the ordinary Hall effect before moving to the magnetic-related part.
\\
\par
From the high-field slope of the Hall effect ($\mu_{0} H$ = \qtyrange{11}{13}{\tesla}) we extracted for each sample the total carrier density using $\rho_{xy} = H/en + a_{0}$, with $e$ -- electron charge and $n$ -- charge density, and $a_{0}$ -- an intercept.
The carrier density determined from the \qty{2}{\kelvin} linear Hall fits are \qtylist{1.2e21; 5.5e20; 6.1e20}{\cm^{-3}} for \qtylist{140; 110; 60}{\nm} flakes, respectively.
These values are very similar to those previously reported for bulk crystals (around \qty{4e20}{\cm^{-3}} \cite{Chen_2020, Li_2021}) and lower than reported for 6-layer thin nanostructures (around \qty{4e21}{\cm^{-3}} \cite{Ma_2026}).
Our sample to sample variation can be explained by considering the impact of intrusive contacts \cite{Gluschke_2020}.
The intrusive contacts artificially draw the current flow to them, which results in a lower measured voltage and, hence, "higher" apparent carrier density.
This is especially important for the \qty{140}{\nm} flake since the contacts are wide and intrude deeply inside the flake.
Moreover, the contact areas in the \qty{140}{\nm} thick flake were etched by \qty{100}{\nm} with an Ar ion beam before the deposition of contacts.
Thus, even more current lines were drawn inside the invasive contacts.
This is also a plausible explanation for the higher values reported in \cite{Ma_2026}. 
Additionally, from the total carrier densities and conductivities, the corresponding averaged mobility values calculated from $\sigma = en\mu$ are $\mu$ = \qtylist{17; 39; 28}{\cm\squared\per\volt\per\s}for \qtylist{140; 110; 60}{\nm} thicknesses, respectively.

To fit the multiband S-shaped Hall effect, a simple two-band model of the form:
\begin{align*}
\rho_{xy} & = \frac{H}{e} \frac{n_{h} \mu_{h}^{2} - n_{e} \mu_{e}^{2} + (n_{h} - n_{e}) (\mu_{h} \mu_{e})^{2} H^{2}}{(n_{h} \mu_{h} + n_{e} \mu_{e})^{2} + (n_{h} - n_{e})^{2} (\mu_{h} \mu_{e})^{2} H^{2}}
\\
\rho_{xx} & = \frac{1}{e} \frac{n_{h} \mu_{h} + n_{e} \mu_{e} + (n_{h} \mu_{e} - n_{e} \mu_{h}) (\mu_{h} \mu_{e}) H^{2}}{(n_{h} \mu_{h} + n_{e} \mu_{e})^{2} + (n_{h} - n_{e})^{2} (\mu_{h} \mu_{e})^{2} H^{2}}
\end{align*}
can be used. Here we assume one single hole and one single electron populations.
The results of the two-band model are presented in SI.
We confirm a majority hole population and a minority electron population, consistent with ARPES data on \ESA.
\\
\par
While the two-band model reproduces very well the $\rho_{xy}$ data above $T_{N}$ = \qty{23}{\kelvin}, a discrepancy is observed at low field below the saturation field of about \qty{5}{\tesla}.
This is apparent on \cref{subfig:AHElin_110nm}, which presents $\Delta \rho^{lin}_{xy} = \rho_{xy} - H/en$ (the same transverse resistivity data as in \cref{subfig:Hall_110nm} after removing the linear high-field Hall slope).
The $\Delta \rho^{lin}_{xy}$ curves measured below $T_{N}$ consist of two S-shape forms, a narrow one between $\pm$\qty{5}{\tesla} and a broader one between $\pm$\qty{14}{\tesla}.
The broader S-shape is present at all temperatures and has a weak temperature dependence.
We interpret this broader S-shape as a consequence of multiband transport, and its weak temperature dependence as an evolution of the carrier mobilities.
The fact that ARPES studies show no band reconfigurations with temperature above and below $T_{N}$ \cite{Li_2019} tends to confirm our interpretation that charge carrier densities are mostly temperature independent.
On the other hand, the inner S-shape only appears below the magnetic transition temperature $T_{N}$.
Moreover, it only develops below the saturation field (see grey dashed line in \cref{subfig:AHElin_110nm}), so that this Hall response is clearly related to magnetism.

To extract the transverse resistivity related to magnetism only, we remove an ordinary Hall background, defined as the \qty{30}{\kelvin} data (above magnetic transition temperature), from the low-temperature Hall data: $\Delta \rho^{30K}_{xy} (H,T) = \rho_{xy}(H,T) - \rho_{xy}(H,T{=}30\mathrm{K})$ (see \cref{subfig:AHE_110nm}).
This choice of background is justified by the fact that the ordinary Hall effect appears mostly temperature independent.
We directly notice that $\Delta \rho^{30K}_{xy}$ does not appear to be proportional to the magnetization curves measured in bulk samples, which are shown in \cref{subfig:AHElin_110nm}.
In order to isolate the residual part from the magnetization-proportional AHE, we present in \cref{subfig:THE_110nm} the residual of $\Delta \rho^{30K}_{xy}$ after removing a component proportional to magnetization: $\Delta \rho^{THE}_{xy}(T) = \Delta \rho^{30K}_{xy}(T) - \alpha \left(M(T){-}M(\mathrm{30K})\right)$, with $\alpha$ a constant.
For each temperature, $\alpha$ was chosen to scale magnetization to the mean value of $\Delta \rho^{30K}_{xy}$ at high-field ($\mu_{0} H>$~\qty{10}{\tesla}).

We observe several features in this residual $\Delta \rho^{THE}_{xy}$.
At low temperature, a pronounced peak is observed below the saturation field, with no hysteresis.
Above $H_{s}$, $\Delta \rho^{THE}_{xy}$ vanishes for all temperatures.
The amplitude of $\Delta \rho^{THE}_{xy}$ decreases with increasing temperature and vanishes around $T_N$.
The fact that $\Delta \rho^{THE}_{xy}$ is non-zero, meaning that there is a Hall contribution other than the ordinary Hall effect or one proportional to the magnetization (anomalous part), is a direct evidence of magnetic scattering on chiral real-space defects appearing below $T_N$ \cite{Kimbell_2022}.
Those textures vanish above $H_{s}$, when spins align along the applied magnetic field, so that no net THE is observed at high field.

We now discuss the origin of $\Delta \rho^{THE}_{xy}$. 
Several chiral magnetic textures could induce such an effect.
First, topological Hall effect was reported in "tripod" magnetic structures (spin chirality fluctuations), where spins rotate around a pole \cite{Cheng_2019, Wang_2019a, Kimbell_2022}.
In \ESA, the reported ground state is A-type antiferromagnetism with an (ab) easy-plane.
A tripod structure in this configuration could possibly emerge if a helical state would develop in the c-axis under an external magnetic field along this direction.
Electrons with a non-zero $k_z$ would feel this helix as an in-plane tripod.
Such an helical structure was reported in the related compound \ch{EuIn2As2} \cite{Takeda_2024, Riberolles_2021} (initially predicted A-type AFM with easy-plane anisotropy) and was reported to induce a topological Hall effect \cite{Yan_2022}.
Other chiral spin fluctuations were also reported to induce a topological Hall effect in the easy-plane A-type antiferromagnet \ch{EuZn2As2} \cite{Yi_2023}.

A second possibility is the existence of uniaxial antiferromagnetic domain walls, which have recently been proposed as a source of a topological Hall effect in \ch{EuAl2Si2} \cite{Xia_2024}.
This is also plausible given the similitude between \ch{EuAl2Si2} and \ESA, both structural (van-der-Waals materials with a magnetic Eu plane in each layer) and magnetic (A-type antiferromagnetic ground state with easy-plane anisotropy).

Lastly, a third possibility would be the existence of magnetic skyrmions, which have been reported as a source of the topological Hall effect \cite{Kimbell_2022}.
A possible cause could be the emergence of a finite Dzyaloshinskii-Moriya interaction in nanostructures due to broken symmetry at interfaces.
Skyrmion lattice and spin Moir\'e superlattices were recently reported in \ch{EuAg4Sb2} \cite{Kurumaji_2025}, a related compound with the same crystallographic structure ($R\bar{3}m$) as \ESA.
Further studies are required to demonstrate which kind of real-space chiral magnetic texture appears in \ESA.

\section{Conclusions}
In conclusion, we performed magnetotransport measurements on exfoliated \ESA\ nanostructures with thicknesses varying between \qtylist{140; 60}{\nm}.
Focusing on the Hall effect, we observe below $T_{N}$ two magnetism-related contributions.
The first contribution is an anomalous Hall effect proportional to the magnetization.
This effect saturates to a non-zero value above the saturation field $H_{s}$.
The second contribution is not proportional to the magnetization.
We understand it as a topological Hall effect, caused by magnetic scattering on real-space chiral magnetic textures.
This topological Hall effect shows a clear peak below $H_{s}$ and disappears above $H_{s}$.
Its amplitude also decreases with increasing temperature, and vanishes around $T_{N}$.
This topological Hall effect could be caused by different types of magnetic textures, such as helical spin textures, antiferromagnetic domain walls or skyrmion lattices, which were all recently reported in similar compounds \cite{Takeda_2024,Riberolles_2021,Xia_2024,Kurumaji_2025}.

Interestingly, similar topological Hall signatures were also observed in other magnetic topological insulators (\MBT\ \cite{Lee_2019} and \ch{EuIn2As2} \cite{Yan_2022}).
Since all those materials have similarities in their crystallographic structures (layered van-der-Waals materials) and magnetic properties (A-type antiferromagnetism), the presence of chiral magnetic textures inducing a topological Hall effect could be a characteristic appearing in a variety of magnetic topological insulators.
This could have consequences in particular for the understanding and tuning of the quantum anomalous Hall effect in these systems.

\begin{acknowledgments}
The authors thank A. Thomas and M. Leroux for discussion about the topological Hall effect.
This work was supported by the Deutsche Forschungsgemeinschaft (DFG, German Research Foundation) project number 449494427.
This work was supported by the CNRS International Research Project “CITRON”.
LV was supported by the Leibniz Association through the Leibniz Competition, and acknowledges support from the French ANR (project number ANR-23-CPJ1-0158-01). 
This work has been partially funded by the Deutsche Forschungsgemeinschaft (DFG) under Project No. 247310070 (SFB 1143), Project No. 390858490 (Würzburg-Dresden Cluster of Excellence on Complexity, Topology and Dynamics in Quantum Matter – ct.qmat, EXC 2147).
\end{acknowledgments}

\section*{Supporting Information}
\setcounter{figure}{0}
\renewcommand{\thefigure}{S\arabic{figure}}

\subsection{Experimental details}

\begin{figure*}
    \centering
    \includegraphics[width=0.8\textwidth]{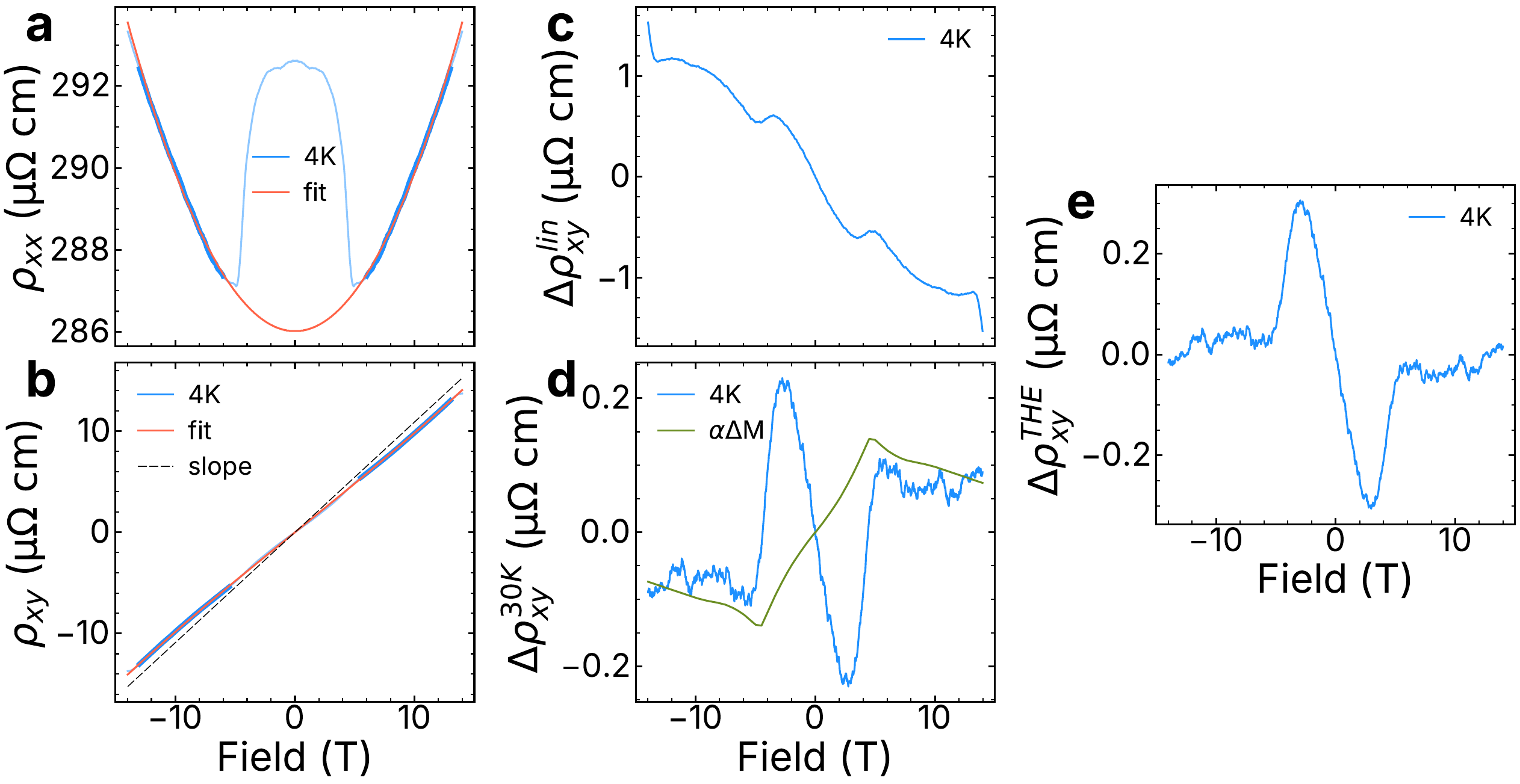}
    \begin{subcaptiongroup}
        \phantomcaption\label{subfig:rhoxx_fit}
        \phantomcaption\label{subfig:rhoxy_fit}
        \phantomcaption\label{subfig:drhoxy_linfit}
        \phantomcaption\label{subfig:drhoxy_30K}
        \phantomcaption\label{subfig:rhoxy_the}
    \end{subcaptiongroup}
    \caption{\label{fig:analysis_procedure} Step-by-step procedure of Hall effect analysis for 110nm flake data at \qty{4}{\kelvin}.
    (\subref{subfig:rhoxx_fit}, \subref{subfig:rhoxy_fit}) Longitudinal $\rho_{xx}$ and transversal $\rho_{xy}$ resistivities along with fit results (dashed line on (\subref{subfig:rhoxy_fit}) denotes line $kB$).
    Thickened portions of the curves indicate the data intervals used for fitting.
    (\subref{subfig:drhoxy_linfit}) Residuals $\Delta\rho^{lin}_{xy} = \rho_{xy} - kB$.
    (\subref{subfig:drhoxy_30K}) Residuals showing magnetic contributions to Hall effect calculated as $\rho^{30K}_{xy} = \rho_{xy}(T)-\rho_{xy}(30\mathrm{K})$ and scaled magnetization $\alpha \Delta M$ ($\Delta M = M(\mathrm{4K})-M(\mathrm{30K})$).
    (\subref{subfig:rhoxy_the}) topological Hall effect $\Delta\rho^{THE}_{xy} = \Delta\rho^{30K}_{xy} - \alpha \Delta M$.}
\end{figure*}

Single-crystals of \ESA\ were grown from a stoichiometric melt cooled down from \qty{850}{\degreeCelsius}; the more detailed description is presented in \cite{Golovchanskiy_2022}.
Flakes were exfoliated from single-crystals using a scotch tape and then pressed onto Si wafers covered with \qty{285}{\nm} thick layer of SiO\textsubscript{2}.
Typical lateral sizes of exfoliated flakes are about \qtyrange{10}{30}{\um} long and \qtyrange{2}{8}{\um} wide.
Despite being a vdW material, \ESA\ exfoliates significantly worse than exemplary representatives of the vdW material family such as graphene or boron nitride. Average thickness of the flakes is about \qty{300}{\nm}, and no color contrast is visible in optical microscopy at least for thicknesses above \qty{30}{\nm} (our thinnest exfoliated flake).
To be able to differentiate thin flakes easier, we covered the substrates with AZ5209E photoresist enhancing the visual contrast.

Contacts were designed in a Hall bar geometry using e-beam lithography (Raith EBL Voyager) or optical lithography (Heidelberg MLA100, for the \qty{140}{\nm} flake).
Then, they were metallized (Ti/Au) in a Plassys e-Beam evaporator and bonded to a sample holder using Al wires.
The studied nanostructures are shown in Fig. 1b of the main text.

Magnetotransport properties were measured in a commercial Quantum Design Dynacool 14T cryostat with a single-axis rotator insert, using standard lock-in techniques.
Current was injected by applying the voltage source to a \qty{1}{\Mohm} polarization resistance in series with the sample studied, with a maximum current \qty{2}{\uA}.

DC magnetic measurements were performed using a Quantum Design Physical Property Measurement System (PPMS) equipped with the Vibrating Sample Magnetometer (VSM) option.
The sample was glued onto a quartz paddle with two quartz shims mounted for out-of-plane measurements.

\subsection{Data processing}

First, we manually removed outliers for longitudinal and Hall data where we considered it is necessary.
This doesn't affect the end results since we removed less than 10 points while data has typically few thousand points.
Next, we interpolated the data to define symmetric common grid and reduce the number of data points.
Then, we smoothed the data using Savitzky-Golay or moving average methods.
Field ranges with edge artifacts due to smoothing have been cropped before presenting the results.
Following this, we symmetrized longitudinal and antisymmetrized Hall data measured in out-of-plane (OOP) configuration to eliminate the effects of contacts misalignment.
As the last step we converted resistance to resistivity as it is more convenient for analysis.

\subsection{Hall data analysis}

To help visualize the analysis we show all involved steps in \cref{fig:analysis_procedure}.
We fitted high-field data with a line $\rho_{xy} = kB + a_{0}$ (\cref{subfig:rhoxy_fit}) and then calculated $\Delta\rho^{lin}_{xy} = \rho_{xy} - kB$ (\cref{subfig:drhoxy_linfit}).
This step serves only the purpose of demonstrating the nonlinear shape of the Hall signal.
Next, we performed self-consistent two-band fitting on cropped $\rho_{xx}$ and $\rho_{xy}$ data.
We removed edge artifacts to account for smoothing.
And we excluded regions of the data affected by magnetic ordering to reduce its influence on the fitted values.
For $\rho_{xy}$, we excluded \qtyrange[range-phrase=\text{ to }]{-5.5}{5.5}{\tesla} range for data measured below \qty{30}{\kelvin}.
For $\rho_{xx}$, we isolated data resembling parabolic branches by excluding arbitrarily selected inner ranges.
Nothing was excluded for the \qty{100}{\kelvin} data. \Cref{subfig:rhoxx_fit,subfig:rhoxy_fit} demonstrate typical regions of data used during the fits together with the fits.

\begin{figure}[!h]
    \centering
    \includegraphics[width=0.48\textwidth]{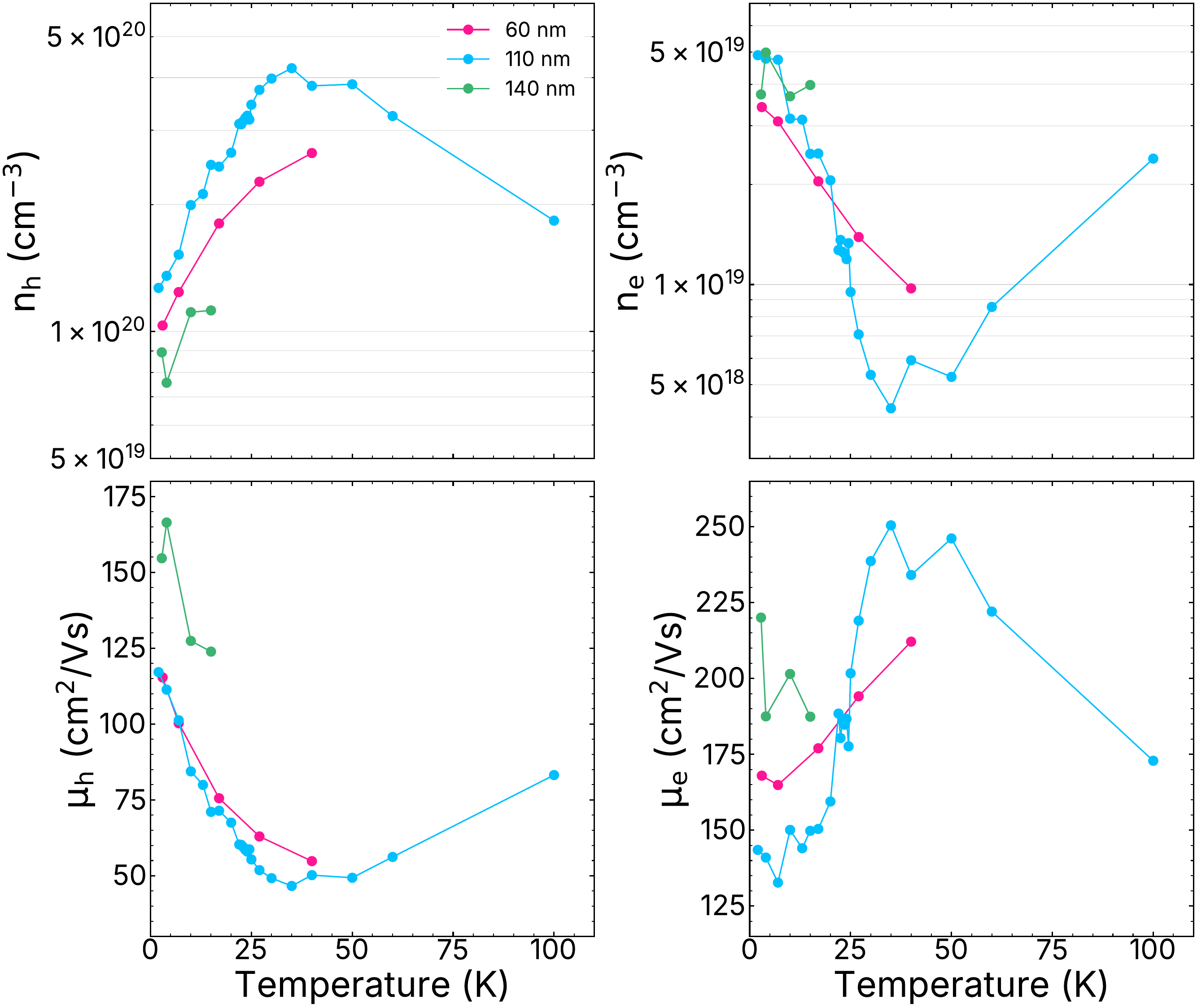}
    \caption{\label{fig:twoband_params} Temperature dependence of free parameters $n_{h}$, $\mu_{h}$, $n_{e}$, $\mu_{e}$ found from two-band fits for \qtylist{140; 110; 60}{\nm} flakes.}
\end{figure}

Temperature dependences of resulting two-band parameters for three samples and temperatures are shown in \cref{fig:twoband_params}.
All four parameters ($n_{1}$, $\mu_{1}$, $n_{2}$, $\mu_{2}$) have been used as free (with \qty{1.6e20}{\cm\tothe{-3}}, \qty{150}{\cm\squared\per\volt\per\s}, \qty{2.5e19}{\cm\tothe{-3}}, and \qty{200}{\cm\squared\per\volt\per\s} as initial vector) during the fit since we had no additional information to fix any.
This results in densities and mobilities of the bands compensating for each other in the optimization problem.
This can be seen in \cref{fig:twoband_params} as density and mobility for each band look like mirror images of each other.
However, values and their temperature trends are comparable between three samples.
Thus, one can expect for real values to be close to our finds. 

\begin{figure}
    \centering
    \includegraphics[width=0.35\textwidth]{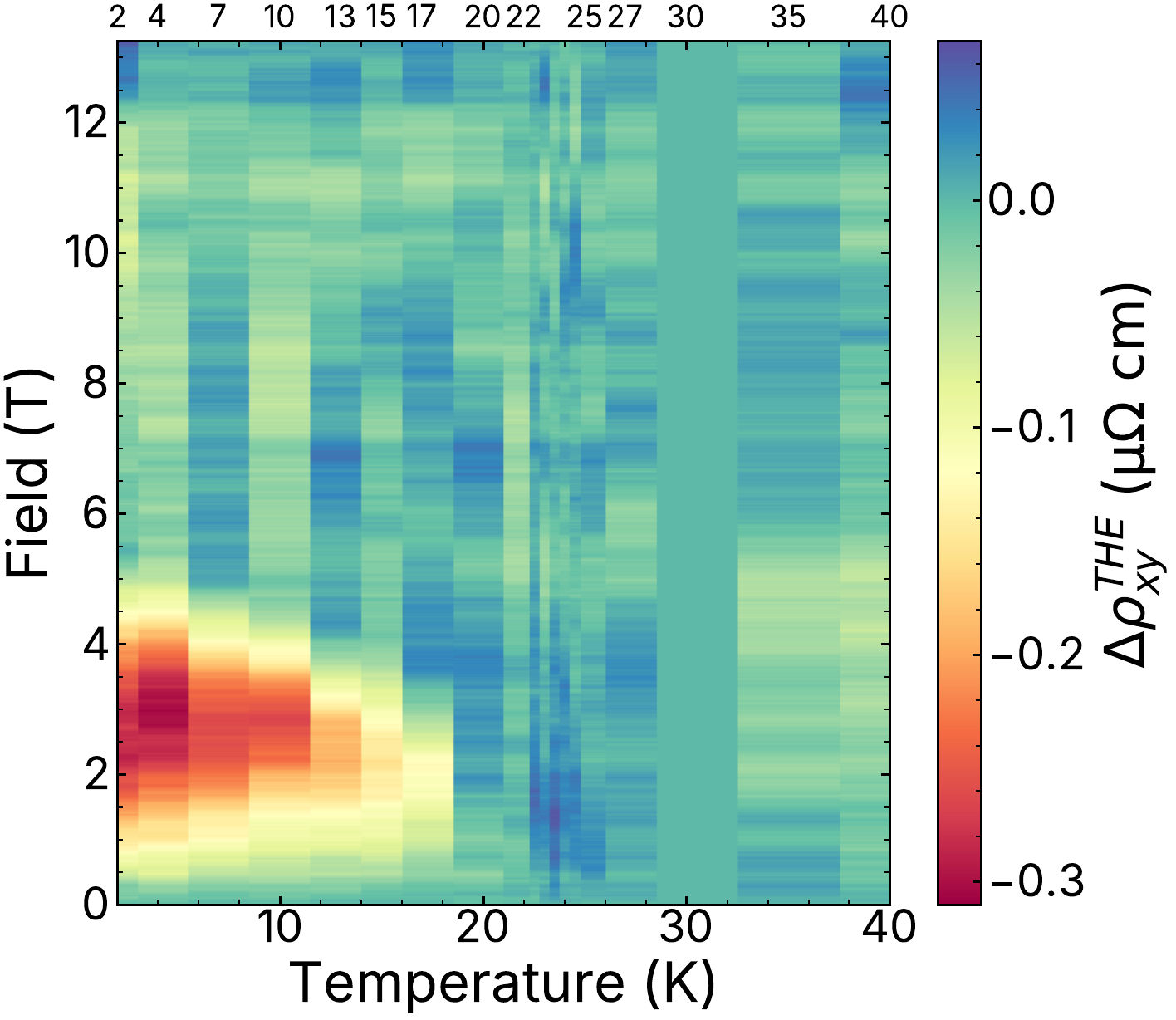}
    \caption{\label{subfig:THE_diagram} Color plot of the topological Hall effect $\Delta \rho_{xy}^{THE}$ in the \qty{110}{\nm} flake} 
\end{figure}

As we argued in the main text, the nonlinearity of the Hall signal comes from three different mechanisms: several charge carrier populations, anomalous Hall effect (AHE), and topological Hall effect (THE). Next step was to separate them.
We subtracted $\rho_{xy} (30\mathrm{K})$ from all other $\rho_{xy}$ curves to remove multi-band nonlinear contribution.
This is done with an assumption that the band-structure doesn't change significantly with temperature which is consistent with temperature-independent Hall curves.
The resulted curves $\Delta\rho^{30K}_{xy} = \rho_{xy}(T)-\rho_{xy}(30\mathrm{K})$ show clear anomalous Hall effect \cref{subfig:drhoxy_30K}.
Next, we interpolated and scaled magnetization measured on a single-crystal from the same batch to the mean value of $\Delta\rho^{30K}_{xy}$ in \qtyrange{10}{14}{\tesla} range (\cref{subfig:drhoxy_30K}).
AHE is generally proportional to the magnetization; so, subtraction of the scaled magnetization $\alpha M(T)$ should remove AHE contribution.
Part of this contribution was already removed together with $\rho_{xy}(30\mathrm{K})$ since magnetization is still significant at \qty{30}{\kelvin}.
We take that into account and define THE as residuals $\Delta\rho^{THE}_{xy}(T) = \rho_{xy}(T)-\rho_{xy}(30\mathrm{K}) - \alpha \left(M(T){-}M(30\mathrm{K})\right)$ (\cref{subfig:rhoxy_the}). In \cref{subfig:THE_diagram}, we show the color plot for $\Delta\rho^{THE}_{xy}$ for temperatures below \qty{40}{\kelvin} (temperature range of the magnetization measurements).

\subsection{Reproducibility}

\begin{figure}
    \centering
    \includegraphics[width=0.35\textwidth]{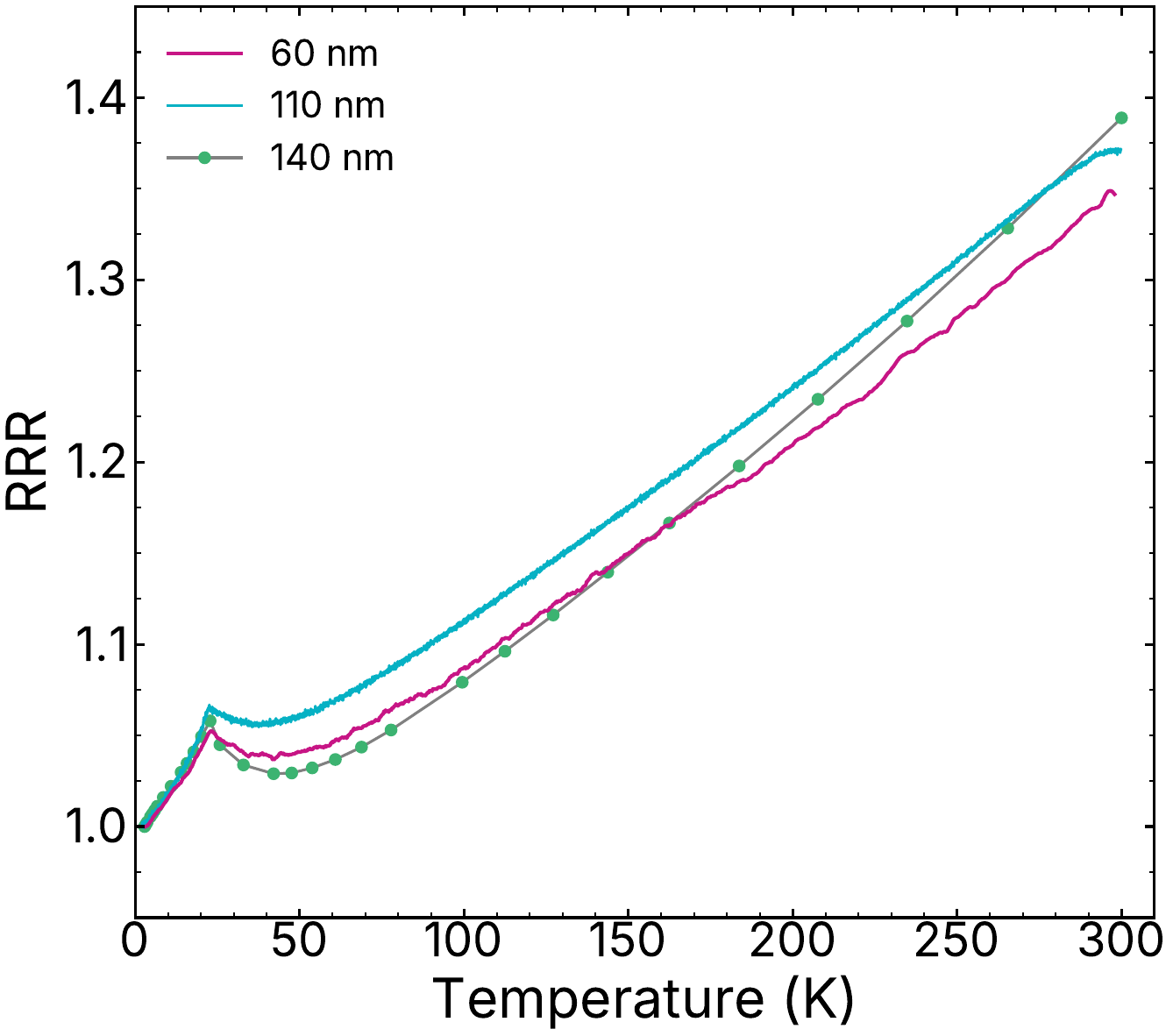}
    \caption{\label{fig:rrr_comparison} Residual-resistance ratio for \qtylist{140; 110; 60}{\nm} flakes.} 
\end{figure}

First, we compare the temperature dependence of the longitudinal resistance.
In \cref{fig:rrr_comparison}, we present residual-resistance ratio (RRR) of the flakes studied. As one can see, all three samples exhibit a magnetic transition at the same temperature, and RRR matches very well below $T_{N}$.
Above transition temperature, samples display a metallic behavior with a small variation in the actual shape. 

\begin{figure*}
    \centering
    \includegraphics[width=\textwidth]{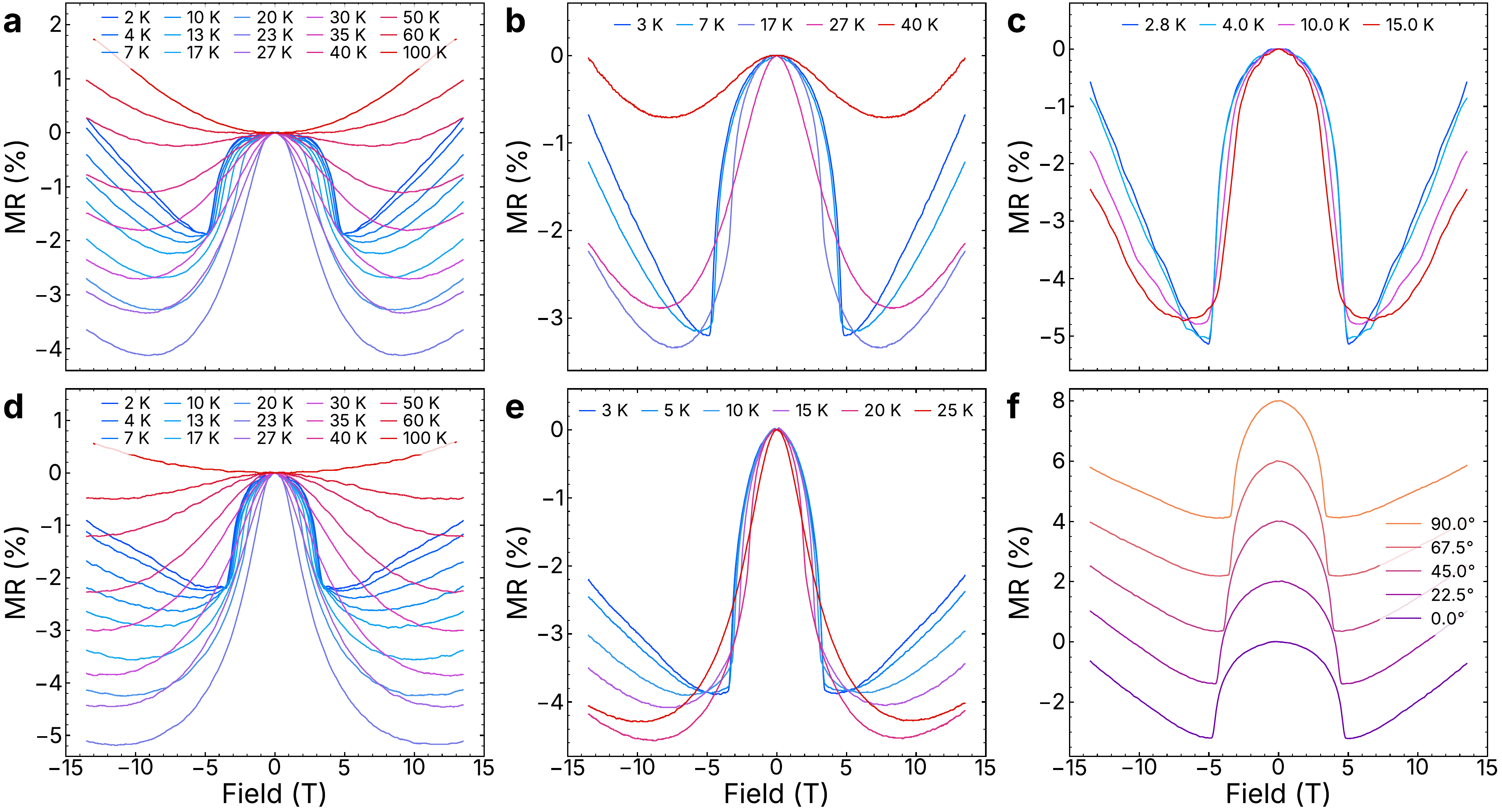}
    \begin{subcaptiongroup}
        \phantomcaption\label{subfig:MR_OOP_110nm_full}
        \phantomcaption\label{subfig:MR_OOP_60nm}
        \phantomcaption\label{subfig:MR_OOP_140nm}
        \phantomcaption\label{subfig:MR_IP_110nm_full}
        \phantomcaption\label{subfig:MR_IP_60nm}
        \phantomcaption\label{subfig:MR_tilt_60nm}
    \end{subcaptiongroup}
    \caption{\label{fig:MR_comparison} Magnetoresistance in \ESA\ flakes.
    (\subref{subfig:MR_OOP_110nm_full}-\subref{subfig:MR_OOP_140nm}) MR at different temperatures in an out-of-plane orientation for \qtylist{110; 60; 140}{\nm} samples.
    (\subref{subfig:MR_IP_110nm_full}, \subref{subfig:MR_IP_60nm}) MR at different temperatures in an in-plane orientation for \qtylist{110; 60}{\nm} samples.
    (\subref{subfig:MR_tilt_60nm}) MR for different field orientations at \qty{2}{\kelvin} for \qty{60}{\nm} sample (vertical shift of \qty{2}{\percent} is added for clarity).
    \ang{0} is out-of-plane, \ang{90} is in-plane.}
\end{figure*}

Second, we can compare magnetoresistance (MR) measured on the flakes in different configurations of the magnetic field (see \cref{fig:MR_comparison}; data for \qty{140}{\nm} was published before). \Cref{subfig:MR_OOP_110nm_full,subfig:MR_OOP_60nm,subfig:MR_OOP_140nm} show MR for \qtylist{110;60;140}{\nm} flakes in out-of-plane configuration at different temperatures.
While the general field dependence is the same, we notice \qtyrange{1}{2}{\percent} of difference in magnitude of the negative MR. \Cref{subfig:MR_IP_110nm_full,subfig:MR_IP_60nm} show MR for \qtylist{110;60}{\nm} flakes in the in-plane (IP) configuration.
The zero-field peak we discussed in the main text for \qty{110}{\nm} flake is not visible on data for \qty{60}{\nm} flake.
However, this is probably caused by the improper alignment of the \qty{60}{\nm} sample and magnetic field in IP case.
While for \qty{110}{\nm} we ensured that current will be parallel to magnetic field, there was an approximately \ang{45} between the current and field for \qty{60}{\nm} sample.
Instead, we observed the presence of the similar peak in raw Hall data which supports this interpretation.
Thus, we report that the zero-field peak is not an artifact of a particular sample but a systematic feature for \ESA.
For our \qty{60}{\nm} flake, we also investigated MR in configurations with tilted magnetic field.
MR displays a similar field dependence for OOP and IP cases indicating a rather isotropic mechanism causing the effect.

\begin{figure*}
    \centering
    \includegraphics[width=0.7\textwidth]{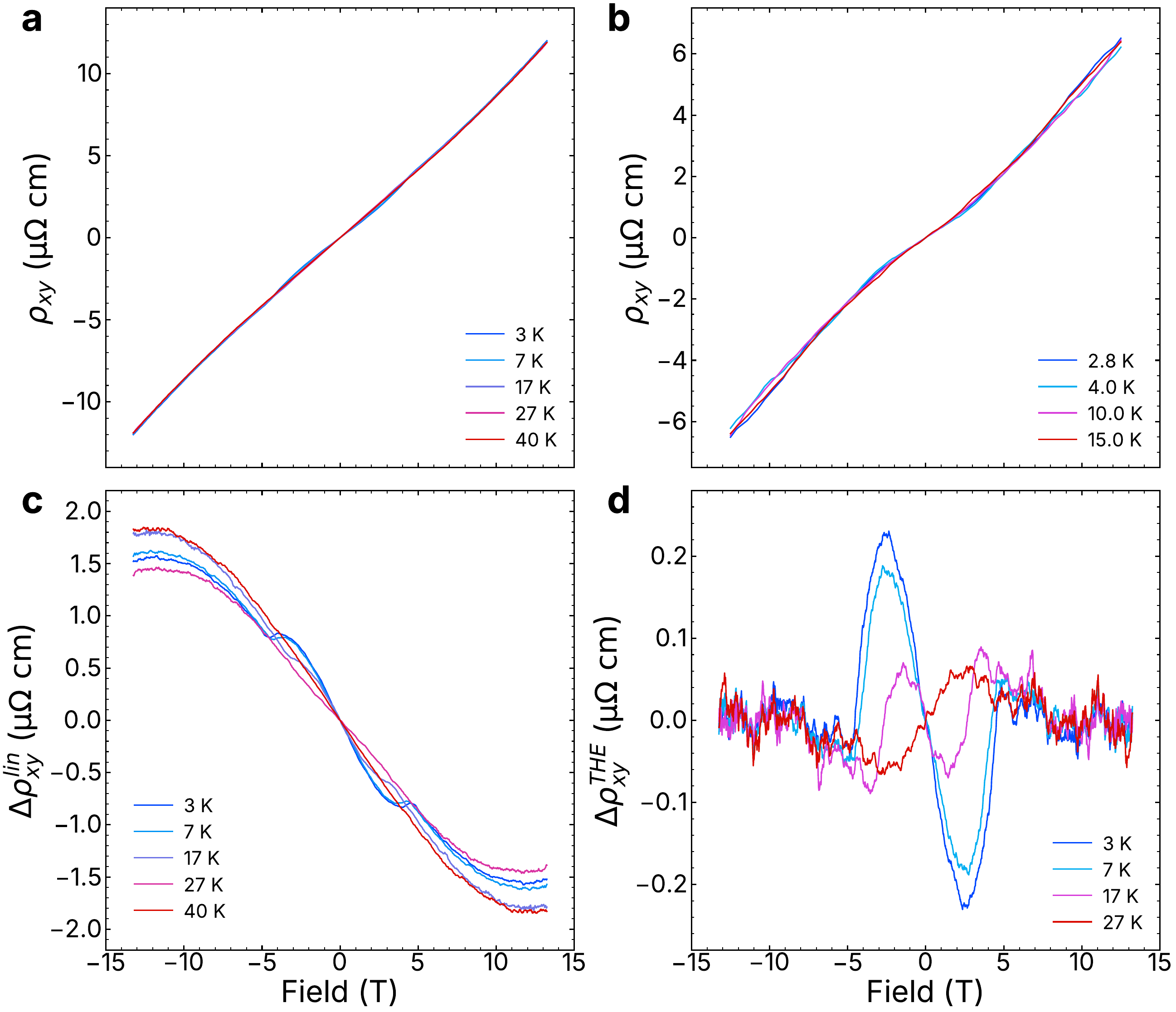}
    \begin{subcaptiongroup}
        \phantomcaption\label{subfig:Hall_60nm}
        \phantomcaption\label{subfig:Hall_140nm}
        \phantomcaption\label{subfig:AHElin_60nm}
        \phantomcaption\label{subfig:THE_60nm}
    \end{subcaptiongroup}
    \caption{\label{fig:Hall_comparison} Hall effect in \ESA\ flakes.
    (\subref{subfig:Hall_60nm},\subref{subfig:Hall_140nm}) Hall effect in \qtylist{60; 140}{\nm} flakes for different temperatures.
    (\subref{subfig:AHElin_60nm}) Residuals of the Hall effect in \qty{60}{\nm} flake after subtracting a high-field (11-13T) linear slope.
    (\subref{subfig:THE_60nm}) Topological Hall effect ($\rho^{THE}_{xy} = \rho_{xy} - \rho^{40K}_{xy} - \alpha \Delta M(T)$) in \qty{60}{\nm} flake.}
\end{figure*}

Next, we provide Hall data measured for \qtylist{60; 140}{\nm} flakes (\cref{subfig:Hall_60nm,subfig:Hall_140nm}; data for \qty{140}{\nm} was published before).
Hall resistivity for \qty{140}{\nm} shows nonlinear field dependence; however, we didn't perform the analysis described above due to the quality of original data.
Although, we analyzed the Hall resistivity for \qty{60}{\nm} flake (\cref{subfig:AHElin_60nm,subfig:THE_60nm}).
For this sample, we investigated the temperature range only up to \qty{40}{\kelvin}.
So, we used $\rho_{xy}(40\mathrm{K})$ as the background to calculated $\Delta\rho^{THE}$.
THE in \qtylist{110; 60}{\nm} flakes matches very well including a comparable magnitude (\qty{0.2}{\micro\ohm\cm}) and evolution with temperature.

\subsection{Zero-field peak}

\begin{figure*}
    \centering
    \includegraphics[width=0.7\textwidth]{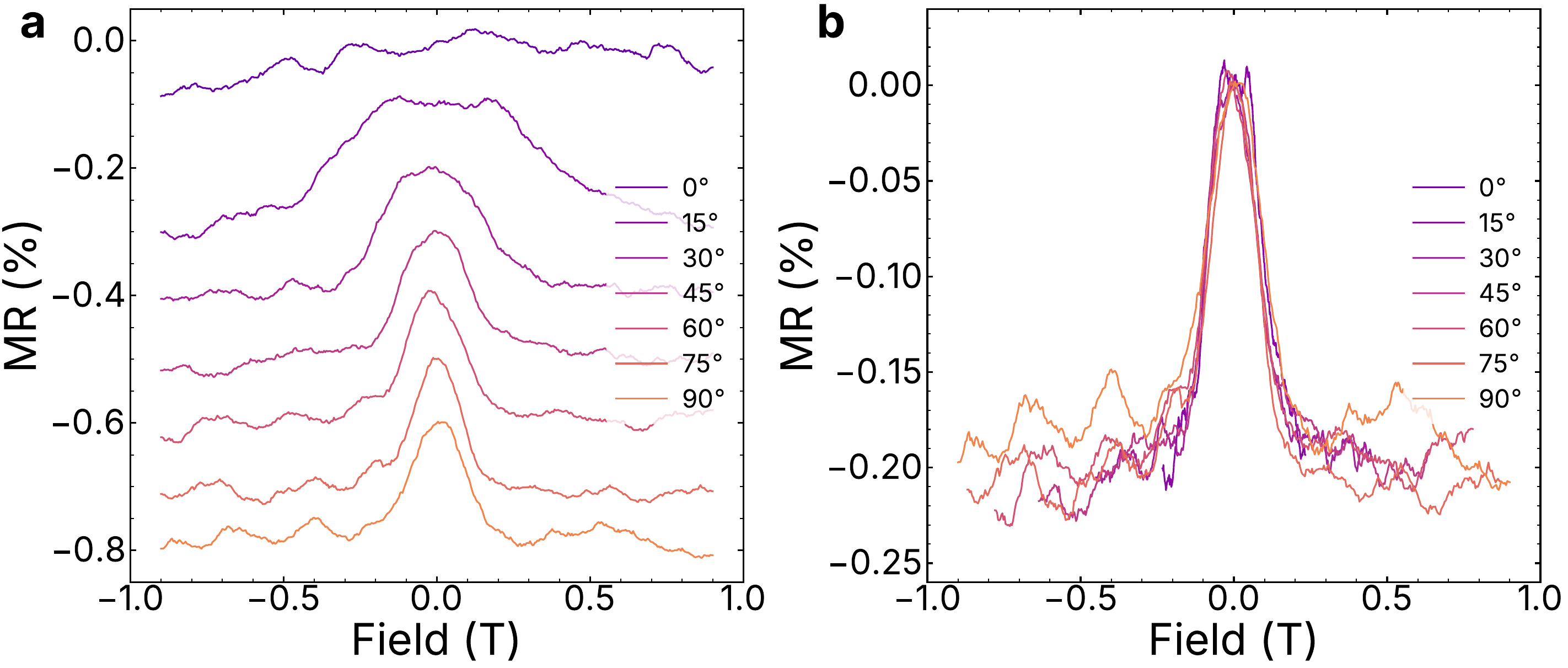}
    \begin{subcaptiongroup}
        \phantomcaption\label{subfig:small_sweeps}
        \phantomcaption\label{subfig:small_sweeps_sin}
    \end{subcaptiongroup}
    \caption{\label{fig:small_sweeps} Magnetoresistance of \ESA\ \qty{110}{\nm} flake for different field orientations at \qty{2}{\kelvin} (data without symmetrizaion).
    \ang{0} is out-of-plane, \ang{90} is in-plane.
    (\subref{subfig:small_sweeps}) Magnetoresistance vs. applied magnetic field $H$ (vertical shift of \qty{-0.1}{\percent} is added for clarity), (\subref{subfig:small_sweeps_sin}) magnetoresistance vs. in-plane component of the applied field $H^{ab} = H \sin(\theta)$} 
\end{figure*}

To investigate the zero-field peak in more detail, we measured magnetoresistance with tilted magnetic field with smaller field steps (\cref{subfig:small_sweeps}).
As the magnetic field tilts towards the in-plane orientation, a peak starts to form.
The peak gradually changes the width with angle which points to the fact that the effect depends only on the in-plane component of the applied field.
To check this, we plotted MR against the in-plane component of the field $H^{ab} = H \sin(\theta)$ (\cref{subfig:small_sweeps_sin}). As one can see all curves collapse to the same shape.

\clearpage

\bibliography{literature}

\end{document}